\documentclass[12pt]{iopart}

\usepackage{iopams}  
\usepackage{color,url,graphicx}

\newcommand\aap{A\&A}%
\newcommand\apj{ApJ}%
\newcommand\apjs{ApJS}%
\newcommand\prd{Phys.~Rev.~D}%
\newcommand\jqsrt{J.~Quant.~Spec.~Radiat.~Transf.}%

\begin{document}

\title[Global Core-Collapse Supernova Comparison]{Global Comparison of Core-Collapse Supernova Simulations in Spherical
Symmetry}

\author{Evan O'Connor$^{1}$,  Robert Bollig$^{2,3}$, Adam
  Burrows$^{4}$, Sean Couch$^{5,6,7,8}$, Tobias Fischer$^{9}$,
  Hans-Thomas Janka$^{2}$, Kei Kotake$^{10}$, Eric J. Lentz$^{11,12}$,
  Matthias Liebend\"orfer$^{13}$, O. E. Bronson Messer$^{14,12,11}$,
  Anthony Mezzacappa$^{15,11}$, Tomoya Takiwaki$^{16}$, David Vartanyan$^{4}$}

\address{
$^{1}$ Department of Astronomy and Oskar Klein Centre, Stockholm University, AlbaNova, SE-106 91 Stockholm, Sweden \\
$^{2}$  Max-Planck-Institut f\"ur Astrophysik, Karl-Schwarzschild-Str. 1, 85748 Garching, Germany \\
$^{3}$ Physik Department, Technische Universit\"at M\"unchen, James-Franck-Stra{\ss}e 1, 85748 Garching, Germany \\
$^{4}$ Department of Astrophysical Sciences, Princeton University, Princeton, NJ 08544 \\
$^{5}$ Department of Physics and Astronomy, Michigan State University, East Lansing, MI 48824, USA\\
$^{6}$ Joint Institute for Nuclear Astrophysics-Center for the Evolution of the Elements, Michigan State University, East Lansing, MI 48824, USA\\
$^{7}$ Department of Computational Mathematics, Science, and Engineering, Michigan State University, East Lansing, MI 48824, USA\\
$^{8}$ National Superconducting Cyclotron Laboratory, Michigan State University, East Lansing, MI 48824, USA\\
$^{9}$ Institute for Theoretical Physics, University of Wroc{\l}aw, 50-204 Wroc{\l}aw, Poland\\
$^{10}$ Department of Applied Physics, Fukuoka University, Nanakuma 8-19-1, Fukuoka 814-0180, Japan\\
$^{11}$ Department of Physics and Astronomy, University of Tennessee, Knoxville, TN 37996-1200, USA\\
$^{12}$ Physics Division, Oak Ridge National Laboratory, P.O. Box 2008, Oak Ridge, TN 37831-6354, USA\\
$^{13}$ Department f\"ur Physik, Universit\"at Basel, CH-4056 Basel, Switzerland\\
$^{14}$ National Center for Computational Sciences, Oak Ridge National Laboratory, P.O. Box 2008, Oak Ridge, TN 37831-6164, USA\\
$^{15}$ Joint Institute for Computational Sciences, Oak Ridge National Laboratory, P.O. Box 2008, Oak Ridge, TN 37831-6354, USA\\
$^{16}$ National Astronomical Observatory of Japan, Mitaka, Tokyo 181-8588, Japan}

\ead{evan.oconnor@astro.su.se}
\vspace{10pt}
\begin{indented}
\item[]June 2018
\end{indented}

\begin{abstract} We present a comparison between several simulation codes designed to study the core-collapse supernova mechanism.  We pay close attention to controlling the initial conditions and input physics in order to ensure a meaningful and informative comparison.  Our goal is three-fold.  First, we aim to demonstrate the current level of agreement between various groups studying the core-collapse supernova central engine. Second, we desire to form a strong basis for future simulation codes and methods to compare to.  Lastly, we want this work to be a stepping stone for future work exploring more complex simulations of core-collapse supernovae, i.e., simulations in multiple dimensions and simulations with modern neutrino and nuclear physics. We compare the early (first $\sim$500\,ms after core bounce) spherically-symmetric evolution of a 20$M_\odot$ progenitor star from six different core-collapse supernovae codes: 3DnSNe-IDSA, AGILE-BOLTZTRAN, FLASH, F{\sc{ornax}}, GR1D, and PROMETHEUS-VERTEX. Given the diversity of neutrino transport and hydrodynamic methods employed, we find excellent agreement in many critical quantities, including the shock radius evolution and the amount of neutrino heating. Our results provide an excellent starting point from which to extend this comparison to higher dimensions and compare the development of hydrodynamic instabilities that are crucial to the supernova explosion mechanism, such as turbulence and convection.  \end{abstract}

%
%
\submitto{\JPG}
%
%
%

\section{Introduction}

Simulations of core-collapse supernovae have a long history, starting in the 1960's with the seminal work of \cite{May_1966,Colgate_1966,Arnett_1966}.  Tremendous progress has been made since then. Today's simulations of core-collapse supernova are incredibly complex. Capturing all of the essential physics requires bringing together input microphysics from nuclear physics, neutrino physics, and stellar evolution, each of which remain uncertain to varying degrees, into multidimensional, general-relativistic, multi-species and multi-energy neutrino-radiation-magnetohydrodynamic simulations. Given the large multi-physics nature of these simulations, the large parameter space of initial conditions, and the varying abilities of individual simulation codes, comparisons between independent investigations have historically been difficult.  

Many comparisons between different neutrino transport schemes have been made in the past.  We briefly summarize some of these comparisons here. The first extensive comparisons were published in \cite{Mezzacappa:1993gx,Mezzacappa:1993gn}, where multi-physics simulations of the infall phase using two codes, one employing Boltzmann neutrino transport and the other employing the multigroup flux-limited diffusion solver of \cite{Bruenn_1985}, were compared across a complete set of hydrodynamic, thermodynamic, and neutrino quantities. In \cite{yamada:99}, a detailed comparison of neutrino transport in static protoneutron star (PNS) atmospheres was done using both Monte Carlo and discrete ordinate Boltzmann transport. There have also been comparisons, both in 1D \cite{Messer_1998} and 2D \cite{Ott_2008}, where the focus was placed on comparing static postbounce snapshots of multi-group flux limited diffusion simulations with the solutions from Boltzmann neutrino transport solvers. \cite{Burrows_2000} compare their variable Eddington factor (with a Boltzmann closure) method to various flux-limiting methods, and recently, \cite{richers:17b} have compared 1D and 2D static configurations using Boltzmann transport and Monte Carlo transport and find excellent agreement. The most extensive comparison to date between two fully independent and dynamic calculations was presented in \cite{Liebendorfer_2005}. This work compared two codes, PROMETHEUS-VERTEX and AGILE-BOLTZTRAN (both are used in this comparison as well). Calculations were done both in Newtonian gravity and general-relativistic gravity, via spherically-symmetric simulations. While AGILE-BOLTZTRAN uses Boltzmann transport via the discrete-ordinate method, PROMETHEUS-VERTEX uses a two moment scheme with a Boltzmann transport-derived closure.  This seminal comparison has been extensively used in the literature as a basis for development of neutrino transport methods.  Recently, \cite{Just_2018} have done a comparison between two codes (PROMETHEUS-VERTEX and AENUS-ALCAR; in both 1D and 2D) using the initial conditions inspired by the work done here.

This article is an attempt to make a fair comparison between many of the core-collapse supernova simulation codes currently in use in the literature. Many comparisons that could be inferred from reviews of the literature, upon closer inspection, are not simulating the same problem. For example, in the two dimensional works of \cite{Bruenn_2016, Summa_2016, O_Connor_2018, Suwa_2015, Vartanyan_2018}, while many of the initial conditions are common, each set of neutrino physics is different in some way. This limits the interpretation of any comparison. Here, in this work, we focus on minimizing differences in the input physics so as to reduce potential sources of disparate results. To this end we use a basic set of neutrino opacities, do not include any nuclear burning, and restrict ourselves to spherical symmetry. In the future, comparisons built off this reference set are encouraged in order to examine more realistic scenarios, including multiple dimensions, nuclear burning, and modern opacities.

This comparison includes the following core-collapse supernova codes, each of which is described in detail below: 3DnSNe-IDSA, AGILE-BOLTZTRAN, FLASH, F{\sc{ornax}}, GR1D, and PROMETHEUS-VERTEX. Overall, we find excellent agreement between all of the codes. We typically see variations in the explored quantities that are $\lesssim$10\%, and in some cases within a few \% across all of the simulation codes. In particular, we see good agreement between highly non-linear quantities, such as the neutrino heating and estimated neutrino detection rates, which depend sensitively on the neutrino spectra being emitted, and in the case of the neutrino heating, the hydrodynamic properties near the supernova shock. In this article, we begin in \S~\ref{sec:setup} by describing the detailed initial conditions, neutrino physics, and non-neutrino physics of our comparison model. In \S~\ref{sec:contribs}, we describe each of the different simulation codes used in this work. We pay special attention to note any differences from the prescribed plan of \S~\ref{sec:setup}. In \S~\ref{sec:results} we show the results of our comparison starting with hydrodynamic quantities, moving on to neutrino quantities and neutrino-matter coupling quantities. We end with comparison of estimated neutrino detection rates in Earth-based detectors. We summarize in \S~\ref{sec:summary}.

\section{The Setup}

\label{sec:setup}
Here we describe, in detail, the initial conditions and the input physics used in this comparison.  We split the description into two main parts, non-neutrino physics and neutrino physics.  Any deviations from the details listed below by specific simulation groups are presented in the following section where the individual codes are described.

\subsection{Non-Neutrino Physics}
We utilize the $20\,M_\odot$ (zero-age main sequence mass), solar metallicity, progenitor from \cite{WOOSLEY_2007} \footnote{We have included this progenitor model as part of the data release associated with this article.  We note that this particular model has the following reference simulation: sollo03/s20/s20\#presn and is also available at \url{http://2sn.org/sollo03/s20@presn.gz}.}. We map the density, temperature, and electron fraction ($Y_e$) from the initial model to each simulation domain. We take the radial extent of the domain to be $10^9$\,cm.  For the sake of clarity, we note that the radial coordinate and the radial velocity in the progenitor model correspond to the value at the outer edge of the zone, while the remaining quantities are zone averages.  We utilize the SFHo equation of state \cite{Steiner_2013}.  This choice is motivated by the large range in density (down to $\sim$1600\,g\,cm$^{-3}$) and temperature (down to 0.1\,MeV) covered by the table that allows us to forgo any additional low-density or low-temperature treatment of the equation of state and thereby removes a potential source of differences between simulation groups. This approach, which assumes nuclear statistical equilibrium (NSE) everywhere, inaccurately captures the composition and therefore the exact equation of state in non-NSE regions of the progenitor star.  However, it ensures consistency between the various simulations. The simulations are performed either with full general relativistic (GR) gravity, or with an effective potential.  For those simulations which utilize an effective potential, \emph{case A} of \cite{Marek2006} is used. For this initial work, we restrict ourselves to spherically symmetric simulations.

\subsection{Neutrino Physics}
For the purposes of this comparison we use a simple, widely implemented, but outdated, set of neutrino opacities. For scattering and absorption on free nucleons we use the rates as presented in \cite{Bruenn_1985}, and also implement weak magnetism and recoil corrections as described in \cite{Horowitz_2002}. For the charged-current absorption rates on free nucleons we do not implement any nucleon potentials other than the neutron-proton rest mass difference.  For scattering on heavy nuclei, we use the \cite{Bruenn_1985} rate, include ion-ion correlations via \cite{Horowitz_1997}, and a correction for the nuclear form factor via \cite{Bruenn_1997} and \cite{Rampp_2002}. For electron-neutrino absorption on nuclei and inelastic neutrino-electron scattering, we implement the rates of \cite{Bruenn_1985}. For pair-processes, we implement both electron-positron annihilation via \cite{Bruenn_1985} and nucleon-nucleon Bremsstrahlung via \cite{Hannestad_1998}.

We use three neutrino species: $\nu_e$, $\bar{\nu}_e$, and $\nu_x = \{\nu_\mu,\ \bar{\nu}_\mu,\ \nu_\tau,\ \mathrm{and}\ \bar{\nu}_\tau\}$. The SFHo EOS contains light clusters ($^2$H, $^3$H, and $^3$He).  While an approximation, all neutrino interactions (both scattering and absorption) on these light clusters are ignored.  We do not reclassify the light clusters as either neutrons and protons, as alpha particles, or as heavy nuclei.

\section{Contributions}
\label{sec:contribs}

In the following we briefly describe each of the simulation codes used in this comparison. We present details of the grid setup, the methods for solving the hydrodynamics, the methods and details of the radiation transport including the energy grid structure and any approximations and/or assumptions made. Each individual group also specifically mentions aspects of their simulation that deviates from the initial conditions and input physics as described in \S~\ref{sec:setup}.

\subsection{3DnSNe-IDSA}
\emph{Contributors: Tomoya Takiwaki, Kei Kotake}

3DnSNe is designed to solve one-(1D), two-(2D), and three-(3D) dimensional hydrodynamics problem in spherical geometry. A piecewise linear method (PLM) with the geometrical correction of the spherical coordinates is used to reconstruct variables at the cell edge, where a modified van Leer limiter is employed to satisfy the condition of Total Variation Diminishing (TVD) \cite{Mignone_2014}.  The numerical flux is calculated by a HLLC solver \cite{Toro_1994}. The computational grid is comprised of 512 logarithmically spaced, radial zones that cover from the center up to the outer boundary of $10^9$\,cm. 
The radial grid is chosen such that the resolution $\Delta r$ is better than 250\,m in the PNS star interior and typically better than 1\,km outside the PNS.  Though the previous works of this code are performed in Newtonian gravity \cite{Takiwaki_2012,Takiwaki_2014,Takiwaki_2016}, the effect of the GR potential is included in this run using Case A of \cite{Marek2006}.

Spectral neutrino transport is solved by the isotropic diffusion source approximation (IDSA) \cite{Liebend_rfer_2009}. While only two species  of neutrinos (electron neutrinos and electron antineutrinos) are included in this scheme in the original version, recently this scheme is extended to treat heavy-lepton neutrinos \cite{kotake_18,Sotani_2016}. In the formalism, the distribution function of the neutrinos is decomposed into a trapped part and a streaming part. The trapped part is once integrated and transported by the hydrodynamic equations. Then its spectrum is reconstructed to satisfy a Fermi-Dirac distribution. The free streaming neutrinos propagate with the characteristic speed following the closure relation \cite{Takiwaki_2014}. In this run, 20 energy groups that logarithmically spread from 1 to 300\,MeV are employed. The velocity-dependent terms ($\mathcal{O}\left(\frac{v}{c}\right)$) are
only included (up to the leading order) in the trapped part of the distribution function (Equation (15) in \cite{Liebend_rfer_2009}). Nucleon-nucleon bremsstrahlung, electron-positron annihilation, and neutrino-electron scattering are included, as described in \cite{kotake_18}. Following \cite{Rampp2002} and \cite{O_Connor_2018}, GR effects (time dilation) are 
approximately taken into account (see Equations (1) -(6) in \cite{kotake_18}). However, gravitational redshifting of the neutrino energies as they leave the gravitational well is not.

\subsection{AGILE-BOLTZTRAN}
\emph{Contributors: Tobias Fischer, Eric Lentz, Matthias Liebend{\"o}rfer, Bronson Messer, Anthony Mezzacappa}
\label{sec:AB}
The radiation-hydrodynamics module AGILE is based on the spherically-symmetric and non-stationary metric of \cite{MisnerSharp:1964} and \cite{MayWhite:1966}. With the choice of orthogonal comoving spacetime coordinates, the equations of hydrodynamics in the presence of the neutrino-radiation field are given in \cite{Liebendoerfer:2001a} and \cite{Liebendoerfer:2004}. They are solved by implicit conservative finite-differencing, with the implementation of a dynamically moving adaptive mass grid following \cite{Winkler:1984} and \cite{DorfiDrury:1987}, allowing for the dynamical allocation of computational zones to regions where they are needed, which ensures an accurate shock capture \cite{Liebendoerfer:2002,Fischer:2009af}. The energy dissipation in the presence of a shock front is considered via artificial viscosity based on the tensor viscosity formalism of \cite{TscharnuterWinkler:1979}. In \cite{Liebendoerfer:2005} AGILE has been upgraded to the second-order total variation diminishing advection scheme based on a Van Leer flux limiter. AGILE employs a flexible equation of state (EOS) module that has been implemented in \cite{Hempel:2012} for the three independent variables temperature, rest-mass density and electron fraction. Contributions from electrons, positrons as well as photons are taken into account following the routines provided by \cite{Timmes:1999} and \cite{TimmesSwesty:2000}.

The shock capturing properties of AGILE are key to its use in the core collapse problem, but the grid reallocation that enables this does have another effect. The motion of zone boundaries in the AGILE scheme produces an effective advection that is essentially second-order in space with a correction term \cite{Liebendoerfer:2002}. The correction term is everywhere present and depends on the grid spacing and the effective grid velocity (with respect to the matter). This induced numerical diffusivity is effective at smoothing initially sharp features in the flow, e.g. discontinuities introduced by burning processes during the late stages of massive star evolution. This effect will be directly responsible for some of the accretion-dependent differences we report in \S~\ref{sec:results}.

The neutrino-transport module BOLTZTRAN consists of a general relativistic time-implicit discrete-angle ($S_N$) multi-species Boltzmann solver. BOLTZTRAN is coupled in an operator-split fashion to the hydrodynamics module AGILE, employing a direct finite-difference representation of the Boltzmann equation \cite{Mezzacappa:1993gm,Mezzacappa:1993gn,MezzacappaMesser:1999,Liebendoerfer:2002,Liebendoerfer:2004}. It solves for the neutrino distribution function, which depends on the spacetime coordinates as well as on the momentum coordinates propagation angle, relative to the radial direction, and neutrino energy. The treatment of inelastic neutrino-lepton scattering has been implemented in \cite{Mezzacappa:1993gx,Fischer:2009}. Neutrinos in specific angle- and energy-bins are created and destroyed according to the collisions. Freely-propagating neutrinos move along light-like geodesics between collisions, which gives rise to many correction terms in the Boltzmann equation due to the use of spherical coordinates in combination with a description of the neutrino phase space in a comoving frame \cite{Liebendoerfer:2004,Liebendoerfer:2005}. The finite difference representation is upward compatible with limiting cases of the Boltzmann equation, e.g., the diffusion limit \cite{Liebendoerfer:2009}, and conserving total energy and lepton number.

The present core-collapse supernova runs are performed with 205 adaptive spatial zones. Solutions of the Boltzmann equation are resolved with 24 energy groups, geometrically increasing following the setup of \cite{Mezzacappa:1993gn}, the first one centered at 0.5\,MeV and the last at 300\,MeV. The propagation angle has been discretized with six bins suitable for Gaussian quadrature.

The AGILE-BOLTZTRAN simulation in this paper deviates from the prescribed plan in the following ways. For the equation of state, the SFHo nuclear EOS is only used above $T\simeq 0.45$\,MeV. Below this temperature, for the present study we consider a low density EOS with only one nucleus $^{28}$Si. There is no nuclear burning of this nucleus to NSE at the transition temperature. In addition, we evolve four ($\nu_e$, $\bar{\nu}_e$, $\nu_x = \{\nu_\mu,\nu_\tau\}$, $\bar{\nu}_x =\{\bar{\nu}_\mu,\bar{\nu}_\tau\}$), instead of the prescribed three, neutrino species.  For the plots below, we average the $\nu_x$ and $\bar{\nu}_x$ values.  Also, we have included neutrino-positron inelastic scattering and extend our domain to $10^{10}$\,cm, though we expect little impact from these particular deviations.

\subsection{FLASH-M1}
\emph{Contributors: Evan O'Connor, Sean Couch}

FLASH \cite{Fryxell_2000,Couch_2014} is an open-source framework for hydrodynamic simulations of astrophysical environments, including core-collapse supernovae.  Recently \cite{O_Connor_2018} have implemented both a general relativistic effective potential \cite{Marek2006} and a two-moment, energy-dependent neutrino transport scheme into FLASH following closely the implementation of \cite{O_Connor_2015} (also see the code description for GR1D below). For this work we use FLASH's unsplit hydrodynamics solver with PPM reconstruction and the hybrid HLLC Riemann solver which reduces to HLLE in the presence of shocks.  Our computational grid uses an adaptive mesh.  We have a total of nine levels of refinement, on the coarsest level we have 160 grid zones extending from the origin to $10^9$\,cm. The grid zones on the finest level are $\sim$244m.  We limit the maximum refinement so as to maintain at least $\Delta r / r \sim 0.009$ resolution. One improvement over \cite{O_Connor_2018} was triggered by this comparison work. We found that triggering mesh refinement based on entropy gradients (in addition to density and pressure) was important to maintain the sharpness of the density gradients near the compositional interfaces. We use 18 neutrino energy groups, spaced logarithmically from 1~MeV to $\sim$275~MeV. For the neutrino transport (also see \S~\ref{sec:GR1D} below), the moment equations are solved in the coordinate frame. We solve the spatial fluxes and the energy-space fluxes explicitly, and the neutrino-matter interactions implicitly.  We retain the full velocity dependence of the moment equations, except for the diffusion limit spatial fluxes, which are approximated to $\mathcal{O}(v/c)$. The explicit flux calculation localizes the solution to each zone (and its neighbors), and avoids expensive matrix solves. 

Our treatment of pair-processes follows that of GR1D (see \S~\ref{sec:GR1D} below). For this comparison we do not include neutrino-electron inelastic scattering. Due to its importance during the collapse phase, we start our FLASH-M1 simulations from 15\,ms after core bounce using a model generated with GR1D (but instead of using full GR we use the general relativistic effective potential to ensure a smooth and self-consistent mapping between the codes).

\subsection{F{\sc{ornax}}}
\emph{Contributors: Adam Burrows, David Vartanyan}

F{\sc{ornax}} \cite{Wallace_2016,Skinner_2016,burrows_2016,Vartanyan_2018,Radice_2017} is a multi-dimensional,  multi-group radiation/hydrodynamic code employing a directionally-unsplit  Godunov-type finite-volume TVD-limited reconstruction method, written  in a covariant/coordinate-independent fashion, with generalized connection  coefficients and static mesh refinement. It solves the comoving-frame,  multi-group, two-moment, velocity-dependent transport equations with an explicit Godunov characteristic method applied to the radiation transport operators and an implicit solver for the radiation source terms, uses the M1 tensor closure for the second and third moments of the radiation fields \cite{Vaytet_2011}, and employs approximate general-relativistic gravity \cite{Marek2006}.

In F{\sc{ornax}}, by addressing the transport operator with an explicit  method, we significantly reduce the computational complexity and communication  overhead of traditional multi-dimensional radiative transfer solutions by  bypassing the need for global iterative solvers that have proven to be slow and/or problematic beyond $\sim$10,000 cores. Radiation quantities are reconstructed with linear profiles, and the calculated edge states are used to determine fluxes via an HLLE solver.  In the non-hyperbolic regime, the HLLE fluxes are corrected to reduce numerical diffusion \cite{O_Connor_2013}.

For these comparison studies in 1D, the F{\sc{ornax}} run employs 16 energy groups for each of three species: $\nu_e$, $\bar{\nu}_e$, and $\nu_{x}$, where the latter subsumes the four known non-electron species.  For the $\nu_e$s, the energy range is 1 to 300\,MeV, spaced logarithmically, and for the other two it is 1 to 100\,MeV, also spaced logarithmically. The radial grid is logarithmic from $\sim$100\,km to the outer boundary and linear interior to $\sim$100\,km, with a central zone width of 0.5\,km. The total number of radial zones is 608. Neutrino sources and sinks due to nucleon-nucleon bremsstrahlung and electron-positron annihilation are included, as described in \cite{Thompson_2000, Burrows_2004, Burrows2006}. Neutrino-electron scattering is based on \cite{Bruenn1985} as implemented in \cite{Thompson_2003}.

We deviate from the prescribed plan as follows, in F{\sc{ornax}}, we use corrections to the neutrino-heavy nuclei scattering cross section based on \cite{Burrows2006} and nucleon-nucleon bremsstrahlung via \cite{Thompson_2000}. Our computational grid extends to $2\times10^9$\,cm.

\subsection{GR1D}
\label{sec:GR1D}
\emph{Contributors: Evan O'Connor}

GR1D \cite{O_Connor_2010,O_Connor_2015} is an open-source, spherically-symmetric core-collapse supernova code.  GR1D is fully general relativistic.  It uses the radial gauge, polar slicing metric of \cite{gourgoulhon:91,Romero_1996,O_Connor_2010}. The hydrodynamics are solved via a second-order Runge-Kutta time stepping with third order spatial reconstruction via the piece-wise parabolic method (PPM).  We couple the neutrinos operator-split from the hydrodynamics.  The neutrino transport is done via an energy dependent M1 scheme where both the zeroth (energy density) and first (momentum density) angular moments of the neutrino distribution function are evolved \cite{Shibata_2011,Cardall_2013,O_Connor_2015}.  The moment evolution equations are closed via an analytic closure that interpolates between the optically thick and optically thin limits of the Eddington factor using the expression from \cite{Minerbo_1978}.  The scheme is fully general relativistic and fully velocity dependent, except in the optically thick limit where the flux of the neutrino moments through cell boundaries is only computed to $\mathcal{O}(v/c)$. The spatial fluxes as well as the energy-space fluxes are computed explicitly, while the neutrino-matter interactions are handled implicitly. The time step is set by the light crossing time of the smallest zone and a Courant--Friedrichs--Lewy (CFL)  factor, which is taken to be 0.5 at all times except for near bounce, when 0.25 is used.  For the simulations presented here, GR1D uses 18 energy groups, spaced logarithmically from 1MeV to 275MeV. The spatial grid uses a constant spaced zoning of 300\,m within the inner 20\,km and outside 20\,km the zoning increases logarithmically until a radius of $\sim 4\times 10^9$\,cm. There are 600 zones in total.

GR1D deviates from the prescribed plan in the following way. Our treatment of pair-processes, like electron-positron annihilation and nucleon-nucleon Bremsstrahlung are treated in an approximate way. First, we do not included thermal processes for electron neutrinos or antineutrinos. Second, for heavy-lepton neutrinos we do not fully solve the non-linear neutrino-matter interaction terms.  Rather, we determine the emissivity of these processes assuming no final state neutrino blocking (via \cite{Burrows_2006}).  We then derive an effective absorption opacity using Kirchoff's law. This approximation is tested in \cite{O_Connor_2015}.  Unlike the prescribed plan, our domain extends to $\sim 4\times 10^9$\,cm.

\subsection{Prometheus-Vertex}
\emph{Contributors: Robert Bollig, Hans-Thomas Janka}

For the integration of the equations of hydrodynamics, PROMETHEUS-VERTEX employs the Newtonian finite-volume hydrodynamics code PROMETHEUS developed by \cite{w1989}. PROMETHEUS is a direct Eulerian, time-explicit implementation of the Piecewise Parabolic Method (PPM) of \cite{Colella_1984}, which is a second-order Godunov scheme based on a Riemann solver. PROMETHEUS is particularly well suited for following discontinuities in the fluid flow like shocks or boundaries between layers of different chemical composition by the help of a contact-steepening technique. Instead of a Newtonian gravitational potential the effective relativistic potential, Case~A, of \cite{Marek2006} is used, adopting the improved energy-conserving implementation of \cite{Muller2010}. The Consistent Multifluid Advection (CMA) method of \cite{1999AampA...342..179P} is applied to ensure accurate advection of the individual chemical components of the fluid.

The VERTEX transport module consists of a time-implicit, conservative integrator of the three-species, energy-dependent moment equations of neutrino energy and momentum. It allows to simultaneously conserve energy and lepton number with good accuracy using the scheme described in \cite{Muller2010}. The neutrino radiation quantities are computed in the comoving frame of the stellar fluid to order $v/c$. Corrections for general relativistic gravitational redshift and time dilation are included. The closure relation for the two-moment set of equations is obtained in the form of a variable Eddington factor (connected to the radiation pressure) and a next-higher-order moment of neutrino radiation intensity, both of which are derived from the solutions of model-Boltzmann equations for all energy bins. The Boltzmann transport equation is simplified with respect to (numerically cumbersome) angular derivatives and integrated in radius-angle space on a tangent-ray mesh. The solutions of the Boltzmann problem and of the two-moment equations are iterated for convergence. Details of this transport code and its coupling to the hydrodynamics solver can be found in \cite{Rampp_2002}
and \cite{Buras_2006}.

The presented stellar collapse simulations were performed with a geometrical grid of 15 energy bins between 0 and 380\,MeV for the boundary of the highest energy bin. The radial grid was contracted with the infalling flow up to core bounce and kept spatially fixed at later times. Initially, the grid contained 400 zones with variable radial spacing chosen such that $\Delta r/r < 0.028$ (except close to the center, where the central zone had a radius of $\sim$0.233\,km). During the simulation the grid was gradually refined in steps such that regions of steep density gradients in the near-surface layer of the proto-neutron star were resolved always with at least 20 radial cells per decade of density.

Unlike the prescribed plan, we have included neutrino-positron inelastic scattering, we do not expect an impact from this.

\section{Results}
\label{sec:results}

In this section we present the results of our comparison. We begin with comparing hydrodynamic quantities and then discuss neutrino related quantities.  Throughout all plots we use the following line style scheme, results from 3DnSNe-IDSA are shown in green, AGILE-BOLTZTRAN results are shown in black, FLASH results are shown in red, Fornax results are in blue, GR1D results are in gray, and PROMETHEUS-VERTEX results are shown in orange.  All results are individually time-shifted so that bounce occurs at $t=0$.  For references, the individual collapse times (time from when the simulation starts to bounce) are $\sim$275\,ms (3DnSNe-IDSA), $\sim$419\,ms (AGILE-BOLTZTRAN), $\sim$299.5\,ms (F{\sc{ornax}}), $\sim$298.2\,ms (GR1D), $\sim$297.8\,ms (PROMETHEUS-VERTEX). All the data presented in this section is available, along with the scripts used to generate the figures. (Please see the Appendix for more information.)

In Figure~\ref{290353} we show the mass accretion rate, $\dot{M} = -4\pi R^2 \rho v$, measured at 500\,km.  Since the accretion onto the shock in the postbounce phase is supersonic, no hydrodynamic information can influence this quantity. Instead, it is mainly effected by the gravitational field, the initial collapse dynamics (before the infall becomes supersonic), and the low density equation of state.  To mitigate differences, all codes, unless otherwise stated above, utilize the same low density equation of state \cite[SFHo;]{Steiner_2013}, map the initial progenitor star via the same prescription (via interpolating density, temperature, and electron fraction), and use general relativistic gravity (or an effective general relativistic gravity). At a postbounce time of $\sim$220\,ms, the silicon-oxygen interface is accreted past 500\,km and results in a steep drop of the mass accretion rate. We show a zoomed view of this region in the inset of Figure~\ref{290353}. This interface, located at a baryonic mass coordinate of $\sim 1.82 M_\odot$, has an initial (at the onset of collapse) density contrast ($[\rho_\mathrm{high}-\rho_\mathrm{low}]/\rho_\mathrm{high})$ of $\sim$40\%. The various hydrodynamic codes maintain the steepness of this density gradient to varying degrees. Regarding the smoothness of the AGILE-BOLTZTRAN result, we remind the reader of the discussion in \S~\ref{sec:AB} regarding the induced numerical diffusivity near sharp features in the flow.   While of small significance in spherical symmetry, this may play a larger role when comparing multidimensional simulations of core-collapse supernovae.
\begin{figure}[h!]
\begin{center}
\includegraphics[width=0.70\columnwidth]{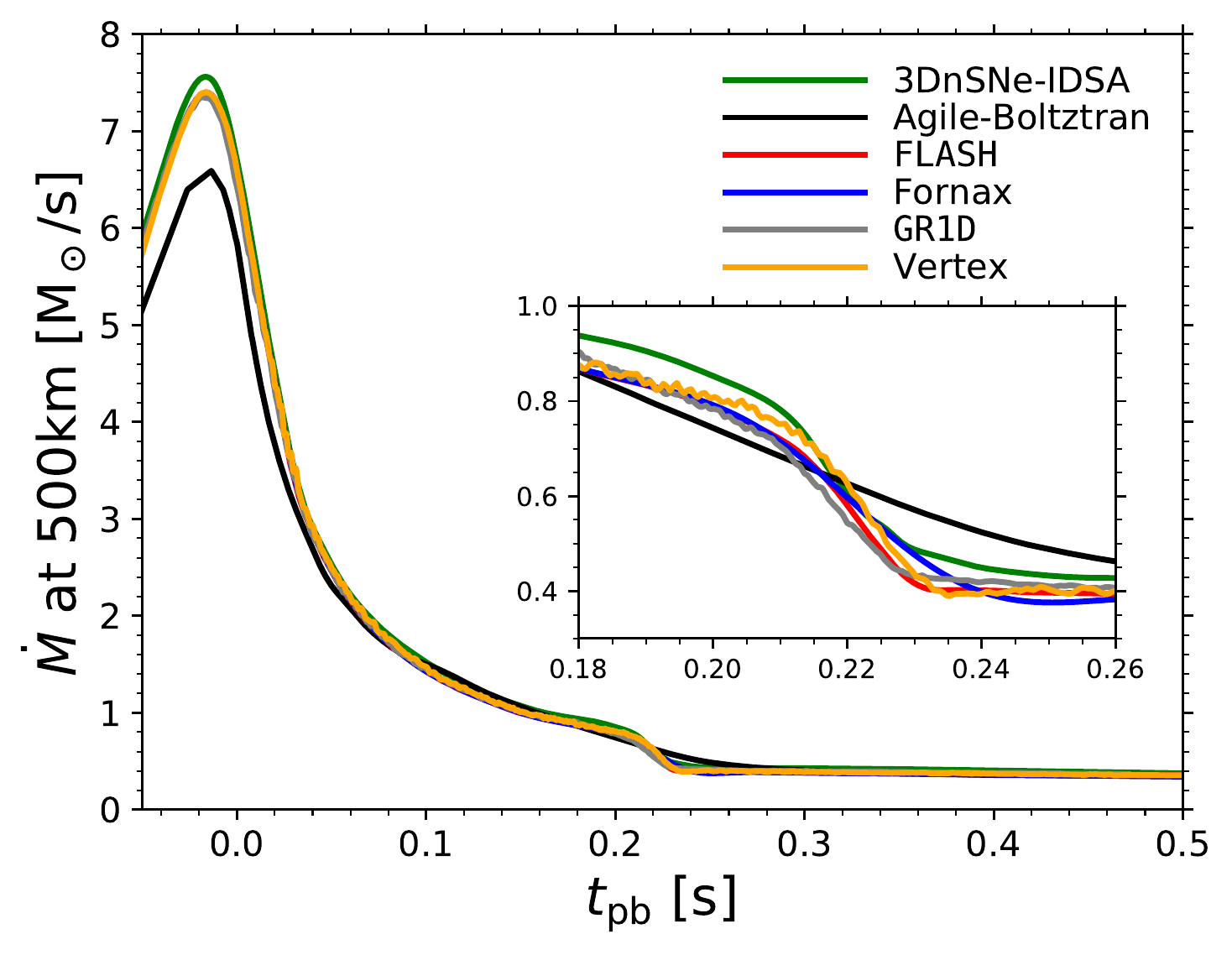}
\caption{{Mass accretion rate measured at 500\,km as a function of postbounce time.
 We show in the inset a zoomed in plot of the mass accretion rate near
the time when the silicon-oxygen interface accretes through 500\,km.
 This gives a significant and steep drop in the mass accretion rate.
 All simulation codes predict a similar postbounce time for this
interface accretion. 
{\label{290353}}%
}}
\end{center}
\end{figure}

\begin{figure}[h!]
\begin{center}
\includegraphics[width=0.70\columnwidth]{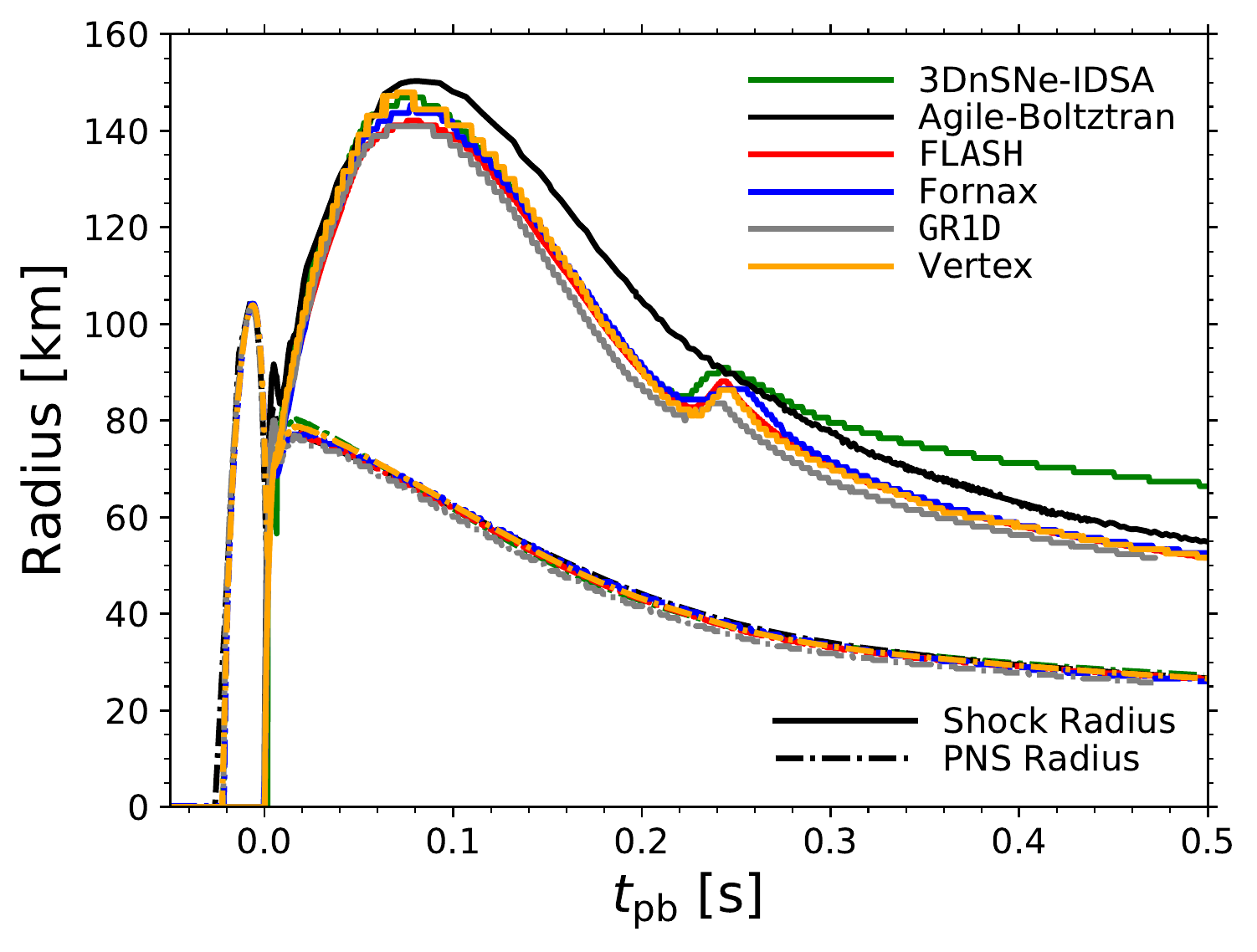}
\caption{{Shock radius (solid) and protoneutron star radius (dashed) evolution as
a function of postbounce time for each simulation in the comparison.
The protoneutron star radius is defined as the radial location with a
density of 10$^{11}$\,g\,cm$^{-3}$, which is why
it is non-zero before bounce, while the shock radius is defined as the
radius where the velocity is maximally negative. In AGILE-BOLTZTRAN the
shock front spans a large radial range, here we take the radius where
the velocity has dropped to half its peak value rather than the radius
of the maximally negative value.
{\label{781038}}%
}}
\end{center}
\end{figure}

The next hydrodynamic quantity we compare is the shock radius.  We show this comparison in Figure~\ref{781038}. The shock forms very close to core bounce and initially travels out into the still infalling iron core.  Due to energy losses from nuclear dissociation and neutrino emission, the shock slows, stalls, and begins to recede.  In spherically-symmetric simulations of this progenitor, we do not expect shock revival.  The codes compare quite well.  The maximum radius reached by the supernova shock is between $\sim$141\,km (GR1D/FLASH) and $\sim$150\,km (AGILE-BOLTZTRAN).  The time when the shock reaches its peak values ranges from $\sim$74\,ms (Vertex) to $\sim$81\,ms (AGILE-BOLTZTRAN). The shock continues to recede until the silicon-oxygen interface reaches the shock.  At this time the mass accretion rate, and therefore the ram pressure, of the material above the shock decreases.  All simulations (except AGILE-BOLTZTRAN) show a transient shock expansion at this time. The extent of the shock expansion depends on the steepness of the drop in the mass accretion rate (see Figure~\ref{290353}). The adaptive Lagrangian grid of AGILE-BOLTZTRAN smooths out the silicon-oxygen interface and erases this feature.  During the late stages, following the accretion of the silicon-oxygen interface we see a range of shock radius recession rates.  

Related to the shock radius is the protoneutron star (PNS) radius. We take, as a definition of the PNS radius, the radius where the matter density $\rho = 10^{11}$\,g\,cm$^{-3}$. We also show this radius in Figure~\ref{781038}, all simulations predict a very similar PNS radius, including before bounce where this quantity simply denotes the radial location of the $\rho=10^{11}$\,g\,cm$^{-3}$ contour.  We note the hierarchy of the PNS radius is generally related to the hierarchy of the shock radius \cite{Janka_2001}.  GR1D has the smallest PNS radius and the smallest shock radius. 

It is worth mentioning that AGILE-BOLTZTRAN and GR1D are the only true GR codes, the remaining codes use an effective potential to mimic GR effects.  Despite this, the agreement across all of the codes, with no obvious systematic offsets between the true GR and effective GR codes, suggests that the GR effective potential does a remarkable job at capturing the relativistic dynamics, at least in spherical symmetry. Nevertheless, one should be cautious when using such an effective potential.  There is one noteworthy, yet subtle, systematic difference between the full GR codes and the effective GR codes that is worth mentioning. Within the first $\sim$5\,ms of core bounce, the behavior of the PNS radius and the shock radius shows an interesting feature in both AGILE-BOLTZTRAN and GR1D that is not present in the codes based on Newtonian hydrodynamics.  Immediately following bounce, there is a faster expansion of these radii, and an additional local maximum that does not occur in the Newtonian runs. This does not appear to impact the subsequent dynamics.

We now focus our attention on comparing neutrino related quantities.  The neutrino field plays a critical role in core-collapse supernovae.  In the neutrino mechanism, the neutrinos are responsible for heating the matter and driving convection,  both of which are crucial for ultimately launching the explosion. The neutrino heating is very sensitive to, and non-linearly depends on, the properties of the neutrino field.  Here we compare the neutrino luminosities, neutrino average energies, and also the total heating in the gain region. With this comparison we hope to show the variations one expects from various neutrino transport methods, the impact on the neutrino heating, and the excellent agreement between the various codes.  Due to the varying definitions of the evolved neutrino variables, both the neutrino luminosities and neutrino average energies are transformed into a frame that is at rest (with respect to infinity). We report the luminosities and energies extracted from a sphere located at 500\,km from the origin. 

\begin{figure*}[h!]
\begin{center}
\includegraphics[width=1.00\columnwidth]{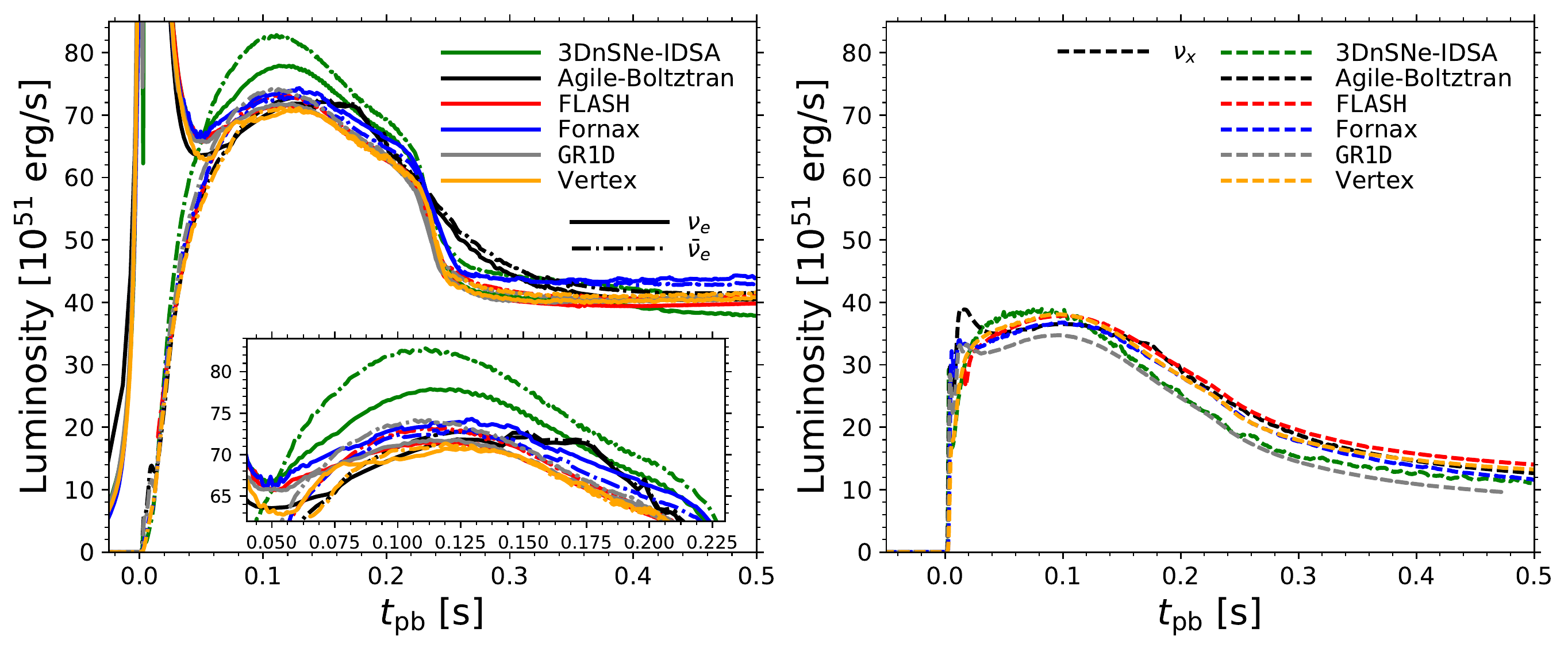}
\caption{{Neutrino luminosities as a function of postbounce time. In the left
panel we show electron-type neutrino luminosities (solid lines show
electron neutrinos while dashed-dotted lines show electron
antineutrinos) and in the right panel we show the characteristic
heavy-lepton neutrino luminosity (dashed line). For clarity, we show an
inset to highlight the early accretion epoch for the electron-type
neutrinos. Some curves have been smoothed with neighboring zones to
remove noise and improve clarity.
{\label{590330}}%
}}
\end{center}
\end{figure*}

In Figure~\ref{590330} we show the electron neutrino (solid lines) and antineutrino (dashed-dotted lines) luminosities in the left panel and the characteristic heavy-lepton neutrino luminosity (dashed lines; for a single species) in the right panel.  For clarity, in the left panel we show an inset of the early accretion phase where the neutrino luminosity plateaus. During this phase, 3DnSNe-IDSA predicts $\sim$10\% higher electron neutrino and antineutrino luminosities when compared to the other codes. These other codes compare well. During the early accretion phase ($\sim$75\,ms - $\sim$200\,ms), the luminosities predicted between these codes vary by at most $\sim$3B/s ($\sim$5\%)  for electron neutrinos and electron antineutrinos. Most of the codes predict a slightly higher $\bar{\nu}_e$ luminosity starting at $\sim$50-75\,ms and continuing through to 500\,ms.

The electron-type luminosities are mainly fueled by accretion, therefore when the accretion rate drops around $\sim$220\,ms, the electron-type luminosities have a corresponding drop.  The roughly constant mass accretion rate following this time is responsible for the flat electron-type luminosities.   After the silicon-oxygen interface accretes and the luminosities plateau again, we find variations of at most $\sim$5\,B/s ($\sim$12\%). As a result of the smoothed mass accretion rate in Figure~\ref{290353} for the AGILE-BOLTZTRAN simulation, the drop at $\sim$220\,ms is not as sharp as the other codes. The heavy-lepton neutrino luminosities show the largest discrepancy among the codes.  The largest absolute difference between any two codes is $\sim$6\,B/s at 400\,ms, which, due to the low absolute luminosity, is upwards of 50\%.  

\begin{figure*}[h!]
\begin{center}
\includegraphics[width=1.00\columnwidth]{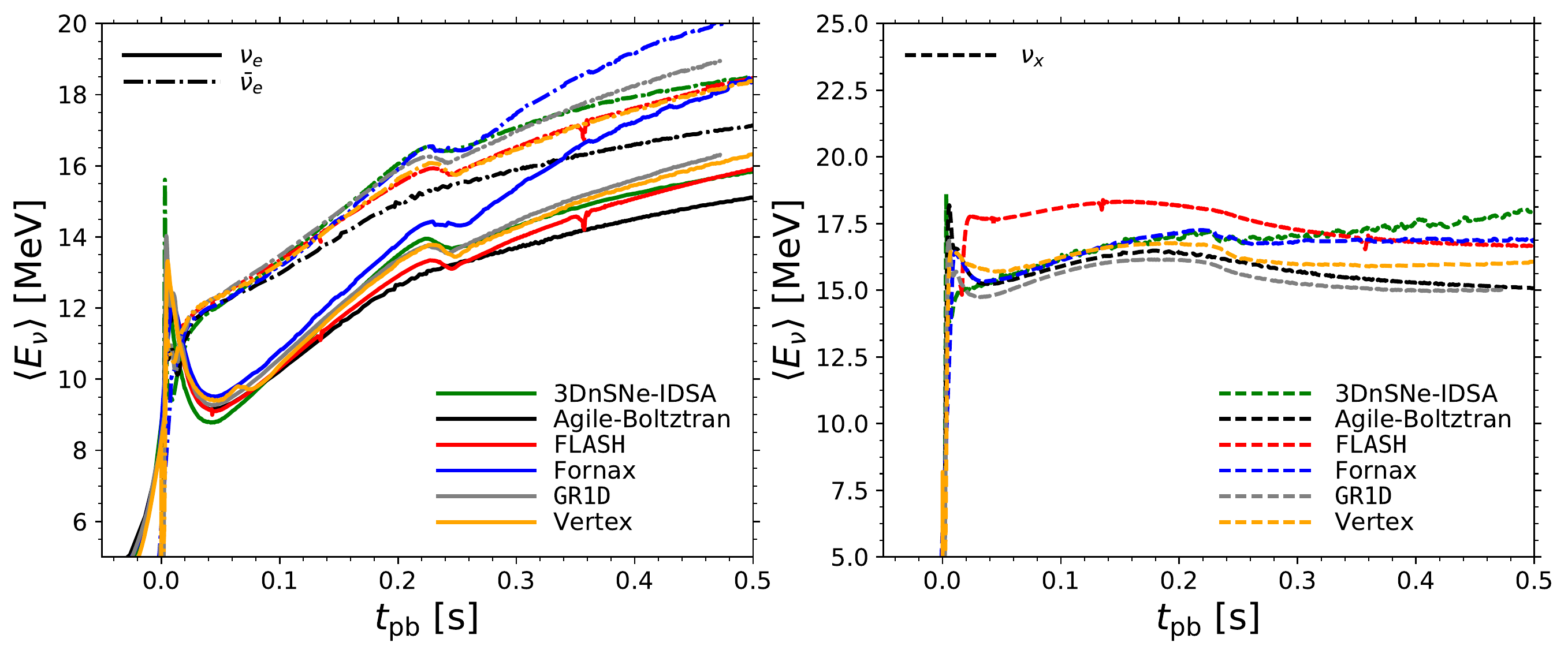}
\caption{{Neutrino average energy as a function of postbounce time. In the left
panel we show electron-type neutrino average energies (solid lines show
electron neutrinos while dashed-dotted lines show electron
antineutrinos) and in the right panel we show the characteristic
heavy-lepton neutrino average energy (dashed line). Some curves have
been smoothed with neighboring zones to remove noise and improve
clarity.
{\label{385693}}%
}}
\end{center}
\end{figure*}

In addition to the neutrino luminosities, we show the neutrino average energies in Figure~\ref{385693}. The average energies are computed by weighting the neutrino energies by the neutrino number spectrum.
 In the left panel we show electron neutrino (solid lines) and electron antineutrino (dashed-dotted lines) average energies while in the right panel we show the characteristic heavy-lepton neutrino average energies (dashed lines). Note, the scales are different. In all simulations we see common features.  The electron neutrino average energies peak at bounce and then reach a minimum around $\sim$~45\,ms after bounce.  They then rise, at a similar rate as the electron antineutrinos, until the silicon-oxygen interface accretes in around $\sim$~220\,ms.  After this, the rise of the mean energies slows.  All codes agree well ($\lesssim$~8\% for electron neutrinos and $\lesssim$~6\% for antineutrinos) until $\sim$200\,ms, after this time we see a divergence.  For the heavy-lepton neutrino energies we see good agreement.  We note that the FLASH $\nu_x$ mean energy is higher, as expected, because neutrino-electron inelastic scattering is omitted in this comparison. Furthermore, the blips in the FLASH mean energies (both electron-type and heavy lepton-type) occur when the shock front passes a mesh-refinement boundary.  At this time, the energy-space coupling terms, which depend on the spatial gradient of the velocity field (which is large at the shock), are adversely impacted by the jump in grid spacing.

Next, we look at the predicted neutrino heating rate in each simulation.  We define this heating as the rate of energy deposition into the internal energy of the matter in zones where this net energy exchange is positive (i.e. neutrino heating in the gain region)\footnote{While included in the simulation, F{\sc{ornax}} does not include the energy exchanged from neutrino-electron scattering in this heating source term. We estimate from the other simulations that this would increase the heating by less than 5\% at the peak and less than 10\% at later times.}. This particular quantity is highly nonlinear in that it sensitively depends on the electron neutrino and antineutrino spectra (both the overall luminosity and also the detailed shape) impinging on the gain region from below as well as the radial structure of the gain region itself and the composition of the matter.  With this understood, we find excellent agreement in the heating rates in all codes.  The rise of neutrino heating begins around 40-50\,ms after bounce.  The peak values, $\sim$~10\,B/s at 100\,ms, are $\lesssim$5\% different from each other.  The heating rate drops after this peak and levels out around $\sim$2-3\,B/s after 250\,ms.
\begin{figure}[h!]
\begin{center}
\includegraphics[width=0.70\columnwidth]{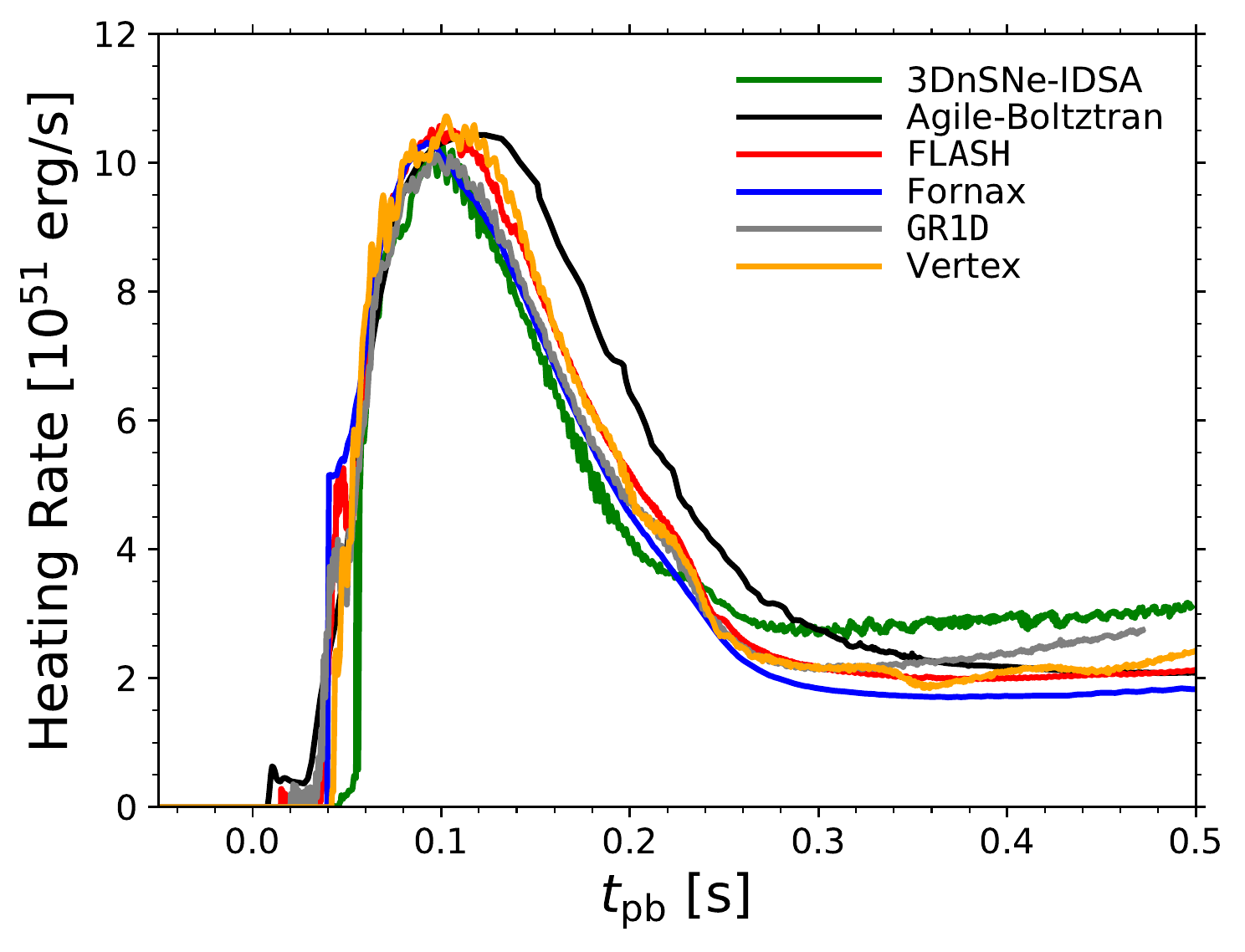}
\caption{{Neutrino heating in the gain region as a function of postbounce time.
We define the neutrino heating to be the change in the internal energy
of the matter due to the interaction with neutrinos.  We only include
contributions to the heating where there is a net transfer of energy to
the matter from the neutrinos. In some simulations we place further
cuts on the data to isolate the gain region, including
$\rho < 3 \times 10^{10}$\,g\,cm$^{-3}$ and $s > 6\,k_{B}$/baryon. Some curves have been smoothed with
neighboring zones to remove noise and improve clarity.
{\label{320544}}%
}}
\end{center}
\end{figure}

\begin{figure}[h!]
\begin{center}
\includegraphics[width=0.70\columnwidth]{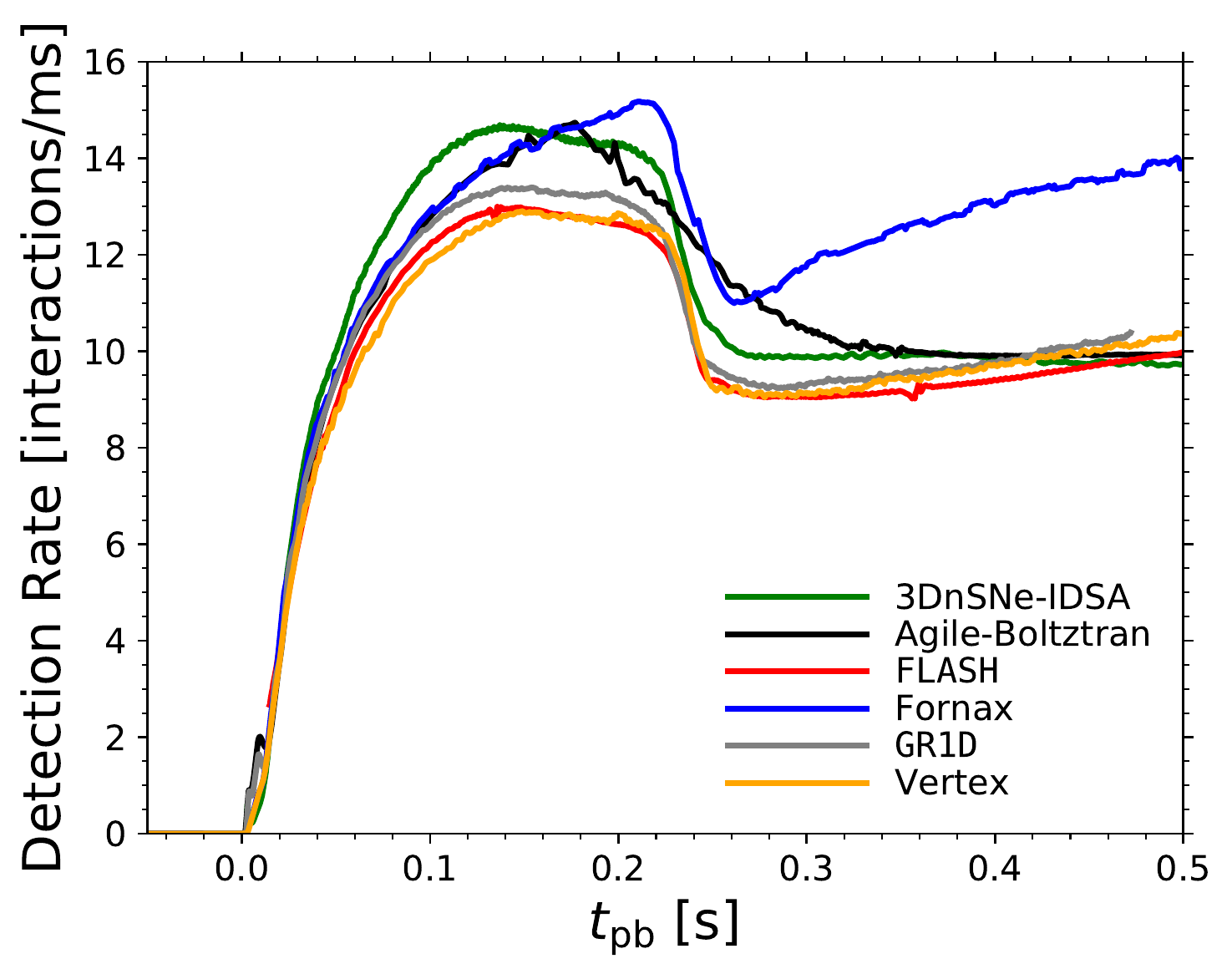}
\caption{{Approximate detection rate in a Super-Kamiokande-like water-Cherenkov
detector for a model supernovae (of a $20\,M_\odot$ progenitor at 10\,kpc based
on Equation~{\ref{eq:detectionrate}}.  This prediction
does not take into account neutrino oscillations. It is meant to
demonstrate a typical systematic uncertainty for neutrino detection rate
predictions.
{\label{421054}}%
}}
\end{center}
\end{figure}

During the next Galactic core-collapse supernovae, many neutrino detectors on Earth will detect neutrinos.  Detailed core-collapse supernova simulations are the only way to predict what this signal will be, and will be critical in aiding neutrino experimentalists and theorists to decipher the detected signal and extract the underlying physics.  As a final comparison, we use the neutrino signals produced by each simulation to determine an approximate rate of neutrino interactions in a Super-Kamiokande-like water-Cherenkov detector due to electron antineutrino capture on free protons from a core-collapse supernova located at a distance of 10\,kpc from Earth from a massive star similar to this 20\,$M_\odot$ model. This prediction is a toy model.  It does not include any neutrino oscillations, or a detailed cross section. Furthermore, it is not complete, it is only for the first 500\,ms, and even then, the impact of removing the spherical symmetry restriction  (for example, potentially allowing an explosion to occur) will significantly alter the signal in reality prior to 500\,ms. Nevertheless, this allows a comparison between the codes. This approximate rate is shown in Equation \ref{421054}.  The interaction rate depends on the number of targets (free protons) in the detector, the inverse beta-decay cross section (electron antineutrino capture on protons), and the electron antineutrino number flux and spectral shape at the detector.  To a good approximation, the inverse beta-decay cross section over the range of energies of interest depends on the neutrino energy squared.  This allows us to write the neutrino interaction rate as a function of the mean squared neutrino energy. We use the following formula for this estimation,

\begin{eqnarray}
\nonumber
R &\sim& \sigma
         \frac{2}{18}\frac{M_\mathrm{det}}{m_\mathrm{amu}}\frac{L_{\bar{\nu}_e}/\langle
         E_{\bar{\nu}_e} \rangle}{4\pi D^2} \frac{\langle
         E^2_{\bar{\nu}_e} \rangle}{(m_e c^2)^2}\\
&\sim&
       \frac{1.6}{\mathrm{ms}}\left[\frac{M_\mathrm{det}}{32\,\mathrm{kT}}\right]\left[\frac{L_{\bar{\nu}_e}}{10^{52}\,\mathrm{erg}\,\mathrm{s}^{-1}}\right]\left[\frac{15\,\mathrm{MeV}}{\langle E_{\bar{\nu}_e} \rangle}\right]\left[\frac{\langle E^2_{\bar{\nu}_e} \rangle^{1/2}}{15\,\mathrm{MeV}}\right]^2\left[\frac{10\,\mathrm{kpc}}{D}\right]^2
\label{eq:detectionrate}
\end{eqnarray}

\noindent
where $\sigma = G_F^2/(\hbar c)^4 \cos(\theta_C)^2 (m_e c^2)^2 (1+3g_A^2) / \pi \sim 2.5 \times 10^{-44}\,\mathrm{cm}^2$, is a reference cross section for absorption of neutrinos onto nucleons \cite{Burrows_2006}; $M_\mathrm{det}$ is the detector mass, here taken to be the inner-volume mass of Super-Kamiokande, 32\,kT \cite{Scholberg_2012}; $m_\mathrm{amu}=1.66054\times10^{-24}$\,g is the atomic mass unit; $D$ is the distance; and $L_{\bar{\nu}_e}$, $\langle E_{\bar{\nu}_e} \rangle$, and $\langle E^2_{\bar{\nu}_e} \rangle$ are the electron antineutrino luminosity, mean energy, and mean squared energy, respectively. The mean energies are taken with respect to the neutrino number spectrum. 

We show the resulting approximate detection rate calculated from each simulation in Figure~\ref{421054}. The initial rise follows the shape of the electron antineutrino signal, it reaches a peak value of $\sim$14 \,interactions/ms after $\sim$100\,ms.  Similar to the neutrino heating, the approximate detection rate is very sensitive to the neutrino spectra.  We see good agreement, especially at early times between all of the simulations.  The largest variation in the detection rate at 100\,ms is $\sim$15\%.  At later times, the deviation in the neutrino energies (and, in particular, also in the mean squared energies; not shown) seen in Figure~\ref{385693} causes a deviation in the approximate detection rate upwards of 40\%.

\section{Discussion and Summary}
\label{sec:summary}

The goal of this work was to bring together groups of researchers studying core-collapse supernovae and collaboratively work together to perform a global comparison between simulation codes.  As we progress forward with multidimensional simulations and begin to successfully model supernovae, it is worth taking the time and effort to convince ourselves, and others, that on a basic and fundamental level, we find broad agreement across independent codes and physics implementations. With this effort we have taken the first steps toward this goal with the first extensive code-to-code comparison in over 10 years, between six core-collapse supernova codes: 3DnSNe-IDSA, AGILE-BOLTZTRAN, FLASH (with its M1 neutrino transport implementation), F{\sc{ornax}}, GR1D, and PROMETHEUS-VERTEX.  We have developed a comprehensive and strict set of initial conditions and input neutrino- and nuclear-physics in order to eliminate as many sources of potential differences between our various simulation codes.  Our goal with this comparison is not to search for and explain any and all differences we find between our various codes (although this has occurred to some extent), or to make statements about which code is more trustworthy than the others. Rather, it is to provide a reference for each other, new researchers to the field, and the external audience on the state of agreement between core-collapse supernova simulation codes. This is in part because in many aspects it is unclear what the correct answer is.  Furthermore, where differences do arise, they may be simply due to various numerical approximations, such as the neutrino transport methods or the hydrodynamics methods, or the numerical implementation of the microphysics. Removing all these differences would be difficult, and even undesirable, as we ideally want to compare the production versions of the various codes. Indeed, it has not been the main goal of the community to drive the convergence of the spherically symmetric case below few percent deviations, which very well could be possible with enough effort. Rather, the ultimate goal in the community is to develop codes that work in multi-D, model core-collapse supernovae as realistically as possible, and show similar convergence.

We have compared select, but critical, aspects from these six codes using spherically-symmetric simulations of the core-collapse and the early post-bounce phase (first 500\,ms) of a core-collapse supernova of a 20\,$M_\odot$ star. These include the mass accretion rate onto the proto-neutron star, the shock radius and protoneutron star radius evolution, the neutrino luminosity and mean energy of each of the three neutrino species included, and the neutrino heating in the gain region.  The mass accretion rates are mostly a test of the low density equation of state and gravity implementations, these agree well between the codes, especially after $\sim$40\,ms.  The shock radius evolution shows good agreement across the codes across the window of time considered, especially, for most codes, during the phase when the interface between the silicon and oxygen shell passes the shock. There we see a transient shock expansion in most codes that all begin within $\sim$10\,ms of each other at a postbounce time of $\sim$225\,ms. With regard to the evolution of the protoneutron star radius during the entire time window, the agreement across all of the codes considered is remarkable.
The neutrino luminosity and mean energy also show good agreement.  During the strongest accretion phase, between $\sim$100\,ms and $\sim$200\,ms, all of the reported electron-type neutrino luminosities (those that drive the neutrino heating) agree within $\sim$15\%, and most agree within $\sim$5\%. The absolute difference in the heavy-lepton neutrino luminosity is similar to the electron-types, but the much lower value gives a larger relative variation, from $\sim$10\% in the accretion phase and growing upwards of $\sim$50\% at 500\,ms. Regarding the neutrino mean energies, at early times they agree well, within $\sim$5\%, but the variation grows with time up to $\sim$25\% at late times. We do note that the evolution of some of our quantities--like the shock radii, neutrino luminosities, and neutrino energies--tend to show some divergence at later times for some of the codes.  We have not diagnosed this difference here, but it useful to keep in mind going forward. The neutrino heating rate is a very non-linear quantity and therefore it is useful to compare it amongst the codes.  It is sensitive to the electron-type neutrino spectra (the overall luminosity, and the spectral shape) as well as the structure of the region just behind the shock front.  We find noteworthy agreement of this quantity in our simulations. The peak heating rate at $\sim$100\,ms after bounce agrees between all the codes to better than $\sim$5\%.  Finally, for each code, we made an approximate prediction for the interaction rate of electron antineutrinos in an Earth-based water-Cherenkov detector similar to the currently-running Super-Kamiokande. While we have left details out of this estimate (like neutrino oscillation effects, detailed cross sections, and detector efficiencies), this estimate is useful for determining a typical systematic error that one can associate with neutrino signal predictions from core-collapse codes.  This error will vary depending on the detector type and the interaction channel, but such a detailed calculation is left to future work. 

In the future, we hope to continue to achieve the agreement seen here while at the same time including more detailed and modern neutrino and nuclear physics, as well as extending our comparison to multiple dimensions.  The growth and impact of multidimensional instabilities, such as convection and turbulence, is likely to depend more sensitively on the choice of hydrodynamic methods and grids.  However, with this base set of comparisons we will have an excellent starting point for this future work.

\section*{Acknowledgments:}

EO would like to thank Almudena Arcones for co-guest editing this Journal of Physics G focus issue and for discussions on the design and goals of this comparison work.

\noindent 
\emph{3DnSNe-IDSA:} KK and TT acknowledge support by the JSPS KAKENHI Grant Numbers
 (JP15H00789, JP15H01039, JP17H01130, JP17H06364, JP17K14306, JP17H05206, and JP18H01212), and 
 by the Central Research Institute of Fukuoka University (Nos.171042, 177103), and by JICFuS 
 which chose the code development of 3DnSNe-IDSA 
as a priority issue to be tackled by using the Post `K' Computer.

\noindent
\emph{Agile-Boltztran:} TF acknowledges support by the Polish National Science Center (NCN) under grant number UMO-2016/23/B/ST2/00720. This work was supported by the COST Actions CA16117 ``ChETEC'' and CA16214 ``PHAROS''. AM acknowledges support from the National Science Foundation (NSF GP 1505933). The Agile-Boltztran supernova simulations were performed at the Wroclaw 
Center for Supercomputing and Networking (WCSS).

\noindent
\emph{FLASH \&  GR1D:} EO acknowledges that partial support for this work was provided by NASA through Hubble Fellowship grant \#51344.001-A awarded by the Space Telescope Science Institute, which is operated by the Association of Universities for Research in Astronomy, Inc., for NASA, under contract NAS 5-26555. SMC is supported by the U.S. Department of Energy, Office of Science, Office of Nuclear Physics, under Award Numbers DE-SC0015904 and DE-SC0017955 and the Chandra X-ray Observatory under grant TM7-18005X. FLASH \& GR1D computations were performed on resources provided by the Swedish National Infrastructure for Computing (SNIC) at the PDC Center for High Performance Computing (PDC).

\noindent
\emph{Fornax}: AB and DV employed computational resources for this study provided by the TIGRESS high performance computer center at Princeton University, which is jointly supported by the Princeton Institute for Computational Science and Engineering (PICSciE) and the Princeton University Office of Information Technology. 
AB acknowledges support under U.S. NSF Grant AST-1714267 and the Max-Planck/Princeton Center (MPPC) for Plasma Physics (NSF PHY-1144374), and by the DOE SciDAC4 Grant DE-SC0018297 (subaward 00009650). 

\noindent
\emph{Prometheus-Vertex:} At Garching, this work was supported by the European Research Council through grant ERC AdG 341157-COCO2CASA, and by the Deutsche Forschungsgemeinschaft (DFG) through the Cluster of Excellence "Universe" (EXC~153) and Sonderforschungsbereich "Neutrinos and Dark Matter in Astro- and Particle Physics" (SFB~1258). Computational resources by the Max Planck Computing and Data Facility (MPCDF) are acknowledged.

\section*{Provided Data}
Upon publication of this article we will provide all the finalized data used to make all of the figures in this article, including the plotting scripts.  This consists of time series data for each simulation code for at least $\dot{M}$ at 500\,km, shock radius, protoneutron star radius, 3 species (electron neutrino, electron antineutrino, and a characteristic heavy-lepton neutrino) neutrino luminosity, mean energy, and mean squared energy (in some cases we provide the pinching parameter $\alpha$, which is analytically related to the mean squared energy), and the net heating rate in the gain region.

\end{document}